\documentclass[a4paper,11pt]{article}
\usepackage{jinstpub} 
\usepackage{subfig}
\usepackage{textcomp}
\usepackage{lineno}
\usepackage{placeins}

\title{\boldmath Production and optical characterisation of blended Polyethylene Terephthalate (PET)/Polyethylene Naphthalate (PEN) scintillator samples}

\author[a,b]{P. Conde Muíño}
\author[c]{J. A. Covas}
\author[a,d]{A. Gomes}
\author[a]{L. Gurriana}
\author[a,b]{R. Machado}
\author[c]{T. Martins}
\author[a]{P. Mendes}
\author[a,b,1]{R. Pedro,\note{Corresponding author}}
\author[a,b]{B. Pereira}
\author[c]{A. J. Pontes}
\author[e]{H. Wilkens}

\affiliation[a]{Laborat\'{o}rio de Instrumenta\c{c}\~{a}o e F\'{i}sica Experimental de Part\'{i}culas - LIP, Lisboa; Portugal}
\affiliation[b]{Departamento de F\'{i}sica, Instituto Superior T\'{e}cnico - IST, Universidade de Lisboa; Portugal}
\affiliation[c]{Instituto de Polímeros e Compósitos - IPC, Universidade do Minho; Portugal}
\affiliation[d]{Faculdade de Ciências - FCUL, Universidade de Lisboa; Portugal}
\affiliation[e]{CERN, Geneva; Switzerland}

\emailAdd{rute@lip.pt, rute.pedro@cern.ch}

\abstract{In Particle and Nuclear Physics research and related applications, organic scintillators provide a cost-effective technology for the detection of ionising radiation. The next generation of experiments in this field is driving fundamental research and development on these materials, demanding improved light yield, radiation hardness, and fast response.
Common materials such as PEN and PET have been found to offer 
scintillation properties competitive to commercial alternatives without the use of dopants. Motivated by their complementarity in terms of light yield, radiation hardness, and time response, there is an increasing interest in investigating PET:PEN mixtures to ascertain whether they exhibit synergistic blending.
This paper presents results from the systematic development of samples of PET, PEN, and PET:PEN blends with varied mass proportions. The manufacturing technique, involving injection moulding of granule raw material, is detailed. The effects of doping the polymer base substrate with fluorescent dopants are explored. Finally, the emission spectra of the different material compositions and their relative light output are presented.}


\arxivnumber{2312.14790} 

\begin{document}
\maketitle
\flushbottom

\section{Introduction}
\label{sec:intro}

Organic scintillators are a cost-effective technology for the detection of ionising radiation~\cite{Birks}, ubiquitously used in Particle and Nuclear Physics research and related applications.
Fundamental R\&D on these materials is being driven by the next-generation experiments in High Energy Physics and their requirements for improved light yield, radiation hardness and fast response~\cite{ECFAroadmap}. Typical plastic scintillators use organic polymeric base solvents, such as Polystyrene or Polyvinyltoluene, doped with wavelength shifters (WLS) in residual mass concentration to bypass the low transmission efficiency of the base materials to the emitted UV light. The addition of WLS fluors also increases the effective scintillation light yield. 

In the 2010s, Polyethylene Terephthalate (PET) and Polyethylene Naphtalate (PEN) have been found to offer scintillation properties competitive with commercial alternatives~\cite{Nakamura:2010,Nakamura:2011} with the advantage of dismissing the introduction of WLS luminophores. The two materials emit blueish light when exposed to ionising radiation (PET emission peaks around 370 to 380~nm~\cite{Nakamura:2010,Nakamura:2011,Nakamura:2013} and PEN at 425 to 480~nm~\cite{Nakamura:2011,Nagata:2013,Nakamura:2013,Majorovits:2017,Efremenko:2019,Kuzniak:2018,Campajola:2022}). Further investigations have shown their complementary characteristics and a PET:PEN blend in 50:50 mass proportion was also explored~\cite{Nakamura:2013}. 
PEN has a superior light yield than PET (10500 against 2200 photons/MeV)~\cite{Nakamura:2011} but the scintillation pulses from PET are faster than those from PEN - the dominant decay constants of PEN and PET were measured to be 35~ns and 7~ns, respectively~\cite{Wetzel:2019,Wetzel:2020,Campajola:2022}. Timing properties are key in experiments with high acquisition rates, for instance at the High Luminosity (HL)-LHC~\cite{HL-LHC} or at the Future Circular hadron Collider (FCC-hh)~\cite{FCChh}. In addition, the large particle fluence foreseen in these experiments leads to scintillator damage and places stringent requirements on the materials used. The radiation hardness of PET and PEN has been probed in a few irradiation tests~\cite{Nagata:2013,Wetzel:2019,Campajola:2021,Campajola:2022}. The light response degradation of PEN was shown to be smaller than that of PET and PEN exhibited a faster recovery. However, PET revealed a larger total recovery after a long period without radiation exposure~\cite{Wetzel:2019}. Besides, the radiopurity of PEN led low background experiments, such as DUNE~\cite{DUNE} or LEGEND~\cite{LEGEND}, to consider this material in their design. Its efficiency in the vacuum-ultraviolet region makes PEN a suitable WLS for the detection of scintillation photons from Liquid Argon-based veto shields~\cite{Kuzniak:2018,Wetzel:2019,protoDUNE} and the mechanical robustness motivated the usage of PEN in active structural components~\cite{Majorovits:2017,Efremenko:2019,Abt:2020,Efremenko:2021}.

The results from the research done so far on the scintillation properties of PET and PEN, reviewed above, reveal complementary attributes of the two materials and motivate the interest in the investigation of PET:PEN mixtures to probe whether they blend synergistically. To answer this question, polymer processing techniques must be adopted to develop samples of PET:PEN blends exploring in a systematic manner different material proportions. 
However, up to now, measurements were done mainly on commercial thin films of pure PET from Goodfellow and thin films of PEN from Teijin Chemicals under the brand names {Teonex\textregistered} and {Scintirex\texttrademark}. The exception was the development and production of structural plates for the LEGEND experiment from white granules/pellets of PEN from Teijin-DuPont~\cite{Abt:2020,Efremenko:2021}. Even the results on PET:PEN blends used samples developed in the industry and do not report on the manufacturing process~\cite{Nakamura:2013}. 

In this paper, we focus on first results from the development of samples of PET, PEN and PET:PEN mixtures with varied mass proportions through injection moulding of materials in pellet form. We further explore the effect of doping the polymer base substrate with fluorescent dopants. The manufacturing procedure is described in Section~\ref{sec:production}.
The characterisation of the scintillation of the produced samples is reported in Section~\ref{sec:characterisation} and presented in terms of emission spectra and relative light response. Characterisation of scintillation timing and radiation hardness will be included in future work.

\section{Production of scintillator samples}
\label{sec:production}

A set of PEN and PET-based scintillator samples were produced at the Institute for Polymers and Composites of the Minho University. The chosen production method was injection moulding, a widely used melt processing technique that allows large productions at low cost.
We injected small scintillator samples ($30\times 30\times 2$ mm$^3$) to test the materials and the production process. The selected base raw materials were PEN granules from Goodfellow~\cite{PENGoodfellow} and amorphous PET granules from SELENIS (PET Selekt BD 110 NATURAL)~\cite{PETSelenis}. 
POPOP~\footnote{1,4-Bis(5-phenyl-2-oxazolyl)benzene} and BBOT~\footnote{2,5-Bis(5-tert-butyl-benzoxazol-2-yl)thiophene} powder from Sigma-Aldrich~\cite{POPOP,BBOT} were chosen to be explored as dopants for PET since they have absorption peaks around 373 and 356~nm respectively, both covering well the emission range of PET. Furthermore, due to the fairly large Stokes’ shifts above 50~nm, they have emission peaks in the blue range, at 407 and 438~nm respectively, well suited for usage with several types of photodetectors.

The raw materials pellets are dried and then mechanically mixed in a container, ready to be fed into the injection moulding machine. 
For doped samples, the dopant is added to the dried pellets and mechanically mixed immediately before injection. The mixed material is fed into the hopper, and melted and homogeneised by a plasticating Archimedes-type screw 
at the selected temperature (305~$^{\circ}$C for PEN and 275~$^{\circ}$C for PET). The processed material is then injected at constant velocity into the cavities of mould and cooled. 
During the initial cooling stages, pressure continues to be applied to the melt, in order to compensate for shrinkage in the cavities.

To avoid humidity absorption
, small amounts of dry raw materials are processed each time. The maximum duration of a production run is 4 hours in the case of mixtures. To avoid contamination from
previous injections, for each type of scintillator, the first 20 samples produced that are already uniform and look good are discarded. 

When using mixtures, for good homogenisation, the materials are left in the plasticating unit at 305~$^{\circ}$C melting temperature for 5 minutes in the machine before injection. 
PET:PEN mixtures were also produced using a small-scale co-rotating twin-screw extruder, but since consistent results could not be obtained, this route was abandoned in the time frame of the project.
In the future, a larger industrial extrusion machine will be considered  for this purpose.

\begin{figure}[t]
\centering
\includegraphics[width=.8\textwidth]{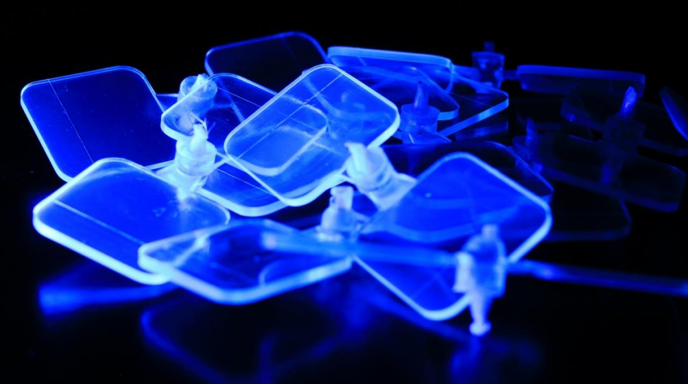}
\caption{PEN samples excited by a UV lantern.\label{fig:PENsamples}}
\end{figure}

The methods described above were used to produce scintillator samples with different material compositions: pure PET and PEN, PET:PEN mixtures with 10:90, 25:75, 50:50, 75:25 and 90:10 mass proportions, PET doped with POPOP and PET doped with BBOT (both in 0.22\% mass concentration). Several samples were produced from each material composition, varying from 15 to 50. 
Figure~\ref{fig:PENsamples} shows a photograph of some of the produced samples excited by a UV lantern, exhibiting the blue light characteristic of the expected scintillation.

\section{Optical characterisation of scintillator samples}
\label{sec:characterisation}

\subsection{Emission Spectra}

\begin{figure}[t]
\begin{center}
\subfloat[PET\label{fig:PETspectra}]{\includegraphics[width=0.5\textwidth]{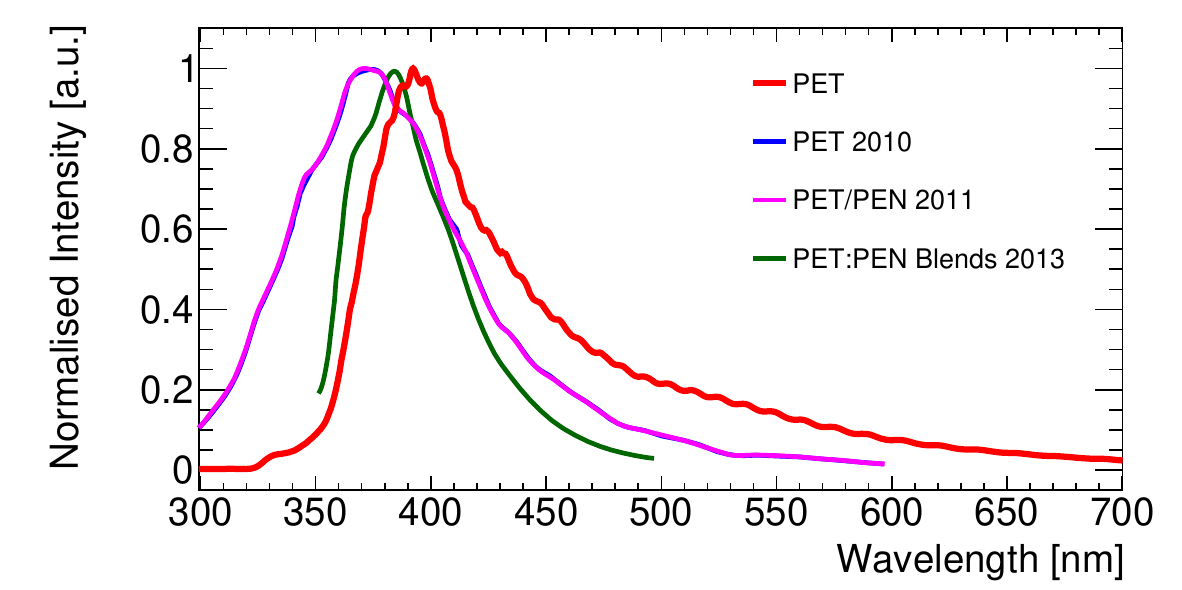}}
\subfloat[PEN\label{fig:PENspectra}]{\includegraphics[width=0.5\textwidth]{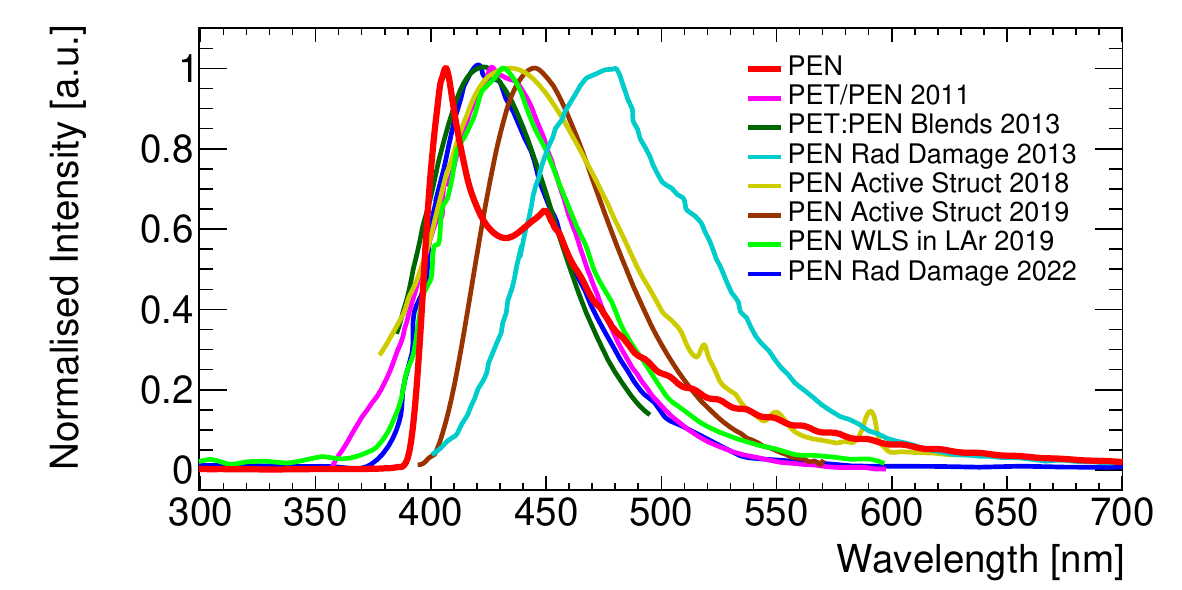}}\\
\subfloat[PET+dopants\label{fig:PETdopants_spectra}]{\includegraphics[width=0.5\textwidth]{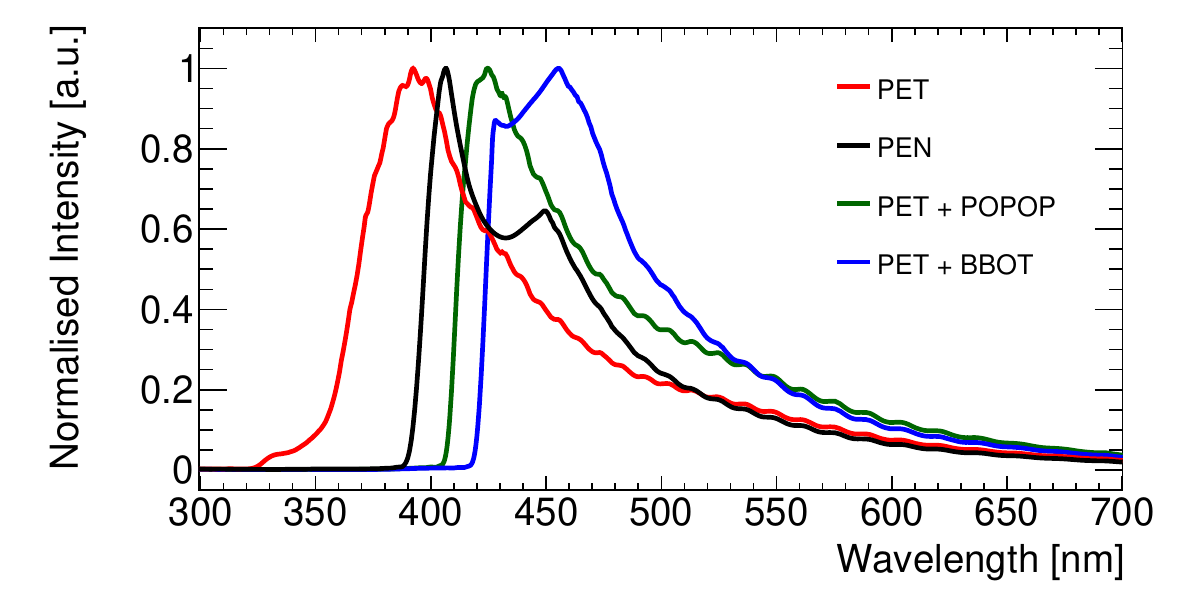}}
\subfloat[PET:PEN mixtures\label{fig:PETPENspectra}]{\includegraphics[width=0.5\textwidth]{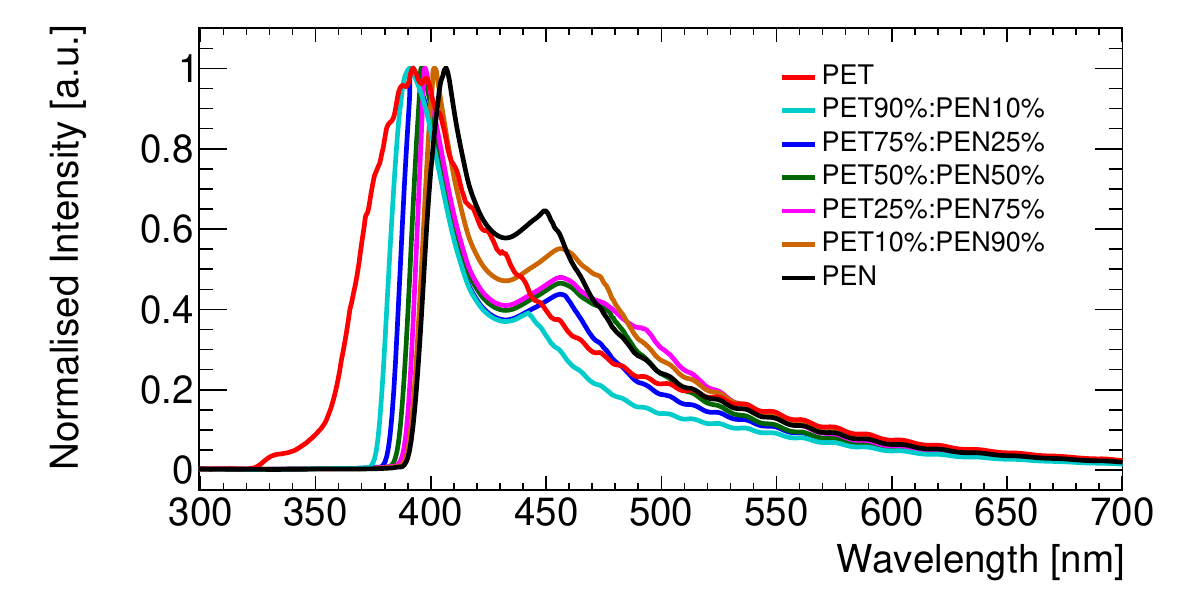}}
\caption{Emission spectra of the (a) PET, (b) PEN, (c) PET+dopants and (d) PEN:PET mixtures samples. Data from the literature is displayed for comparison (PET 2010~\cite{Nakamura:2010}, PET/PEN 2011~\cite{Nakamura:2011}, PET:PEN blends 2013~\cite{Nakamura:2013}, PEN Rad. Damage 2013~\cite{Nagata:2013}, PEN Active Struct. 
2018~\cite{Majorovits:2017}, PEN Active Struct.
2019~\cite{Efremenko:2019}, PEN WLS in LAr 
2019~\cite{Kuzniak:2018}, PEN Rad. Damage 2022~\cite{Campajola:2022}). 
\label{fig:spectra}}
\end{center}
\end{figure}

The emission spectra of the scintillator samples are measured with a ILT960UVIR  spectrometer from International Light Technologies (spectral sensitivity range from 230 to 1050~nm and 2.3~nm resolution)~\cite{ILT960UVIR} using a UV LED with an emission peak between 270 to 280~nm~\cite{LED-UV}, chosen to excite the lower tail of the PET emission spectra (expected to be between 300~nm to 500~nm). The results are shown in Figure~\ref{fig:spectra}.

Figures~\ref{fig:PETspectra} and~\ref{fig:PENspectra} contain the measured spectra of the PET and PEN pure samples, respectively. Data from similar measurements in the literature are superimposed for comparison~\cite{Nakamura:2010,Nakamura:2011,Nakamura:2013,Nagata:2013,Majorovits:2017,Kuzniak:2018,Efremenko:2019,Campajola:2022}. For the PET sample, the peak of the emission spectrum (around 390~nm) is similar to those published (peaks at 370 and 380~nm)~\cite{Nakamura:2010,Nakamura:2011,Nakamura:2013}, differing only about 2 to 5\%. The emission ranges varies in different publications, 300--500~nm for ~\cite{Nakamura:2010,Nakamura:2011} and 360-470~nm for~\cite{Nakamura:2013}, while the PET sample obtained in this work emits in the range between 360--600~nm.
The produced PEN samples emit in the 390 to 600~nm wavelength range, matching the known emission region of the material. The main emission peak is located at around 410~nm, slightly below the emission peaks previously published, which range from 425 to 480~nm~\cite{Nakamura:2011,Nakamura:2013,Nagata:2013,Majorovits:2017,Kuzniak:2018,Efremenko:2019,Campajola:2022}. The samples exhibit a second well-defined emission peak around 450~nm. This is not visible in the published spectra and could be attributed to differences in the source material composition: while our samples are made of PEN granule from Goodfellow~\cite{PENGoodfellow}, the data in the literature refers to {Teonex\textregistered} PEN from Teijin-Dupont either in the form or thin films or injection moulded plates. We highlight that the double peak structure does not imply a disadvantage for the generality of scintillator applications.


The emission spectra of the doped PET samples and PET:PEN mixtures are presented in Figures~\ref{fig:PETdopants_spectra} and~\ref{fig:PETPENspectra}. The spectra of PET and PEN pure samples from Figures~\ref{fig:PETspectra} and~\ref{fig:PENspectra} are also drawn for reference. The addition of POPOP and BBOT to the PET base substrate causes the wavelength shift of the original scintillation light and the resulting peaks are around 425 and 460~nm, both larger than PEN. 

The spectra of the PET:PEN mixtures depend on the mass proportion. The emission peak gradually shifts from 390~nm (pure PET peak) to 410~nm (pure PEN main peak) with increasing PEN proportion, as expected. The second emission peak of the PEN material appears in the mixed samples with a relative intensity that also correlates with PEN concentration. Moreover, the effect of PEN as ultraviolet WLS, reported in~\cite{Kuzniak:2018,Wetzel:2019}, is visible through the suppression of the pure PET near-ultraviolet scintillation (330 to 380~nm) in mixtures with as low as 10\% mass concentration of PEN.

\begin{figure}[t]
\centering
\includegraphics[width=.7\textwidth]{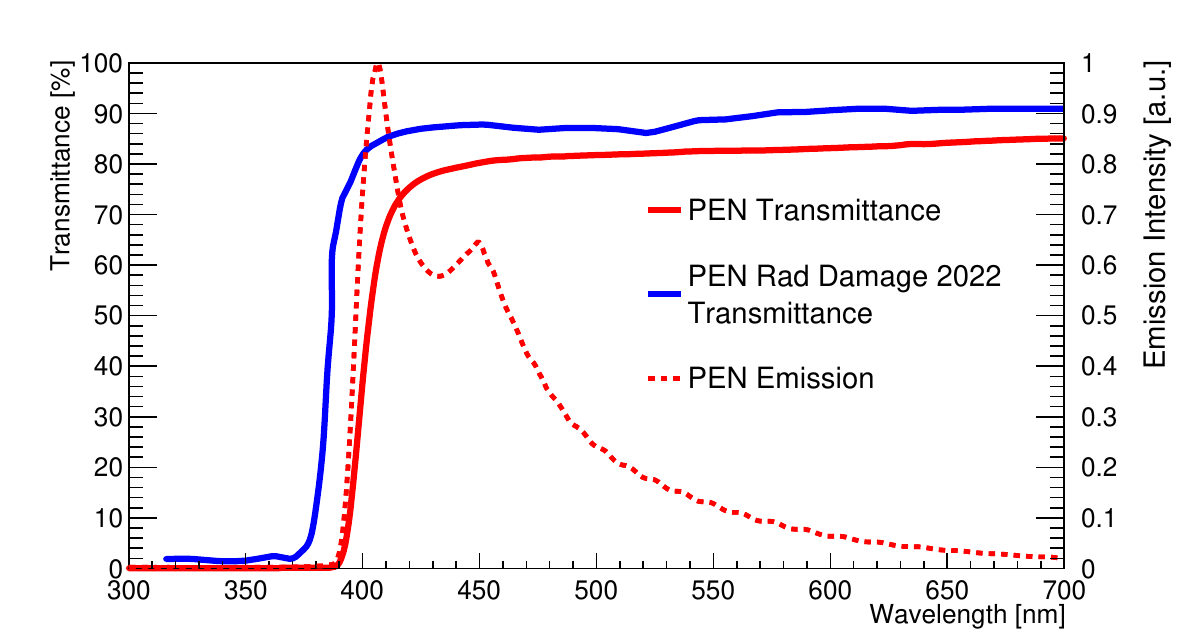}
\caption{Emission and transmittance spectra of a PEN sample. Data from the literature on the PEN transmission spectrum is displayed for comparison (PEN Rad. Damage 2022~\cite{Campajola:2022}).
\label{fig:PENtransmittance}}
\end{figure}

A spectrophotometer (Shimadzu UV-1280)~\cite{Shimadzu-UV-1280} is used to measure the transmittance spectrum of a pure PEN sample, shown in Figure~\ref{fig:PENtransmittance}. The transmittance spectrum of a 100~$\mu$m thick PEN film~\cite{Campajola:2022} is also displayed for comparison. Both measurements are done with the light beam crossing the samples in the transversal direction. The results show good agreement, with the difference in the largest transmittance region, of 80\% for the PEN sample and 90\% for the thin film, being attributed to the different scintillator thickness (2~mm for the sample and 100~$\mu$m for the film).
PEN exhibits transparency above 400~nm and the turn-on of the transmission spectrum matches the lower wavelength region of the emission spectrum, which is superimposed for comparison. This is a strong indication that scintillation below 400~nm is attenuated by the transmission characteristics of the material, pointing to a possible improvement on the light response of PEN by adding an adequate WLS.

\subsection{Light Response}

The experimental setup used for the relative light response measurements consists of an $XY$-movable $^{90}$Sr radioactive source, a readout photomultiplier tube (Hamamatsu H13175U-110~\cite{Hamamatsu-H13175U-110} with spectral response of 230 to 700~nm) and a digital multimeter (Schlumberger SOLARTRON 7150), which integrates the signal over 400~ms. The photomultiplier is supplied by a $-800$~V high voltage using a Canberra 3002 HV power supply. The acquisition and control of the system, including the step motor adjusting the $^{90}$Sr source position, is established by LabVIEW. Two scintillators are measured: a 3~mm thick scintillator from the ATLAS/LHC Tile Calorimeter production (tile Nr. $4$, trapezoidal shape  127~mm high and 262 to 249.5~mm wide)~\cite{TileCalOptics,TileScintillators}, used as a reference for the measurement, and the manufactured scintillating sample to measure. The radioactive source is collimated and produces a $\sim$3 mm diameter spot in the test sample. The scintillation light is collected from one side of each scintillator with independent WLS optical fibres (1~mm diameter double-clad Y11(200)MSJ from Kuraray~\cite{David:2002kva}) which transport it to a common photomultiplier readout.  Both scintillators are wrapped in a Tyvek{\sffamily\textregistered} envelope to improve light collection by the fibres. The two fibres are 100~cm long and were chosen to have similar light collection efficiency and attenuation length to provide similar transmission efficiency. 

\begin{figure}[t]
\centering
\includegraphics[width=.8\textwidth]{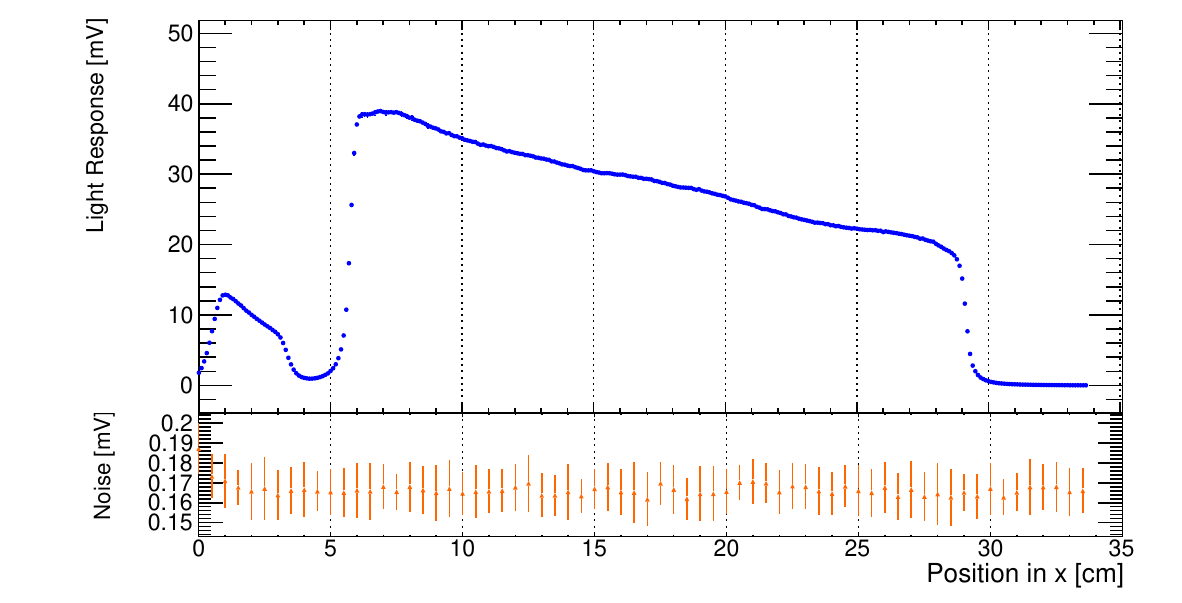}
\caption{Light response (in mV) as a function of $^{90}$Sr source position of the PEN sample ($x\in [0.7,3.7]$~cm) and the reference scintillator plate ($x\in [6,28.5]$~cm). The readout WLS fibres are located around $x=0.6$~cm (PEN sample) and $x=5.9$~cm (reference scintillator).
\label{fig:PENscan}}
\end{figure}

The two scintillators are scanned from the bottom along the direction transverse to the fibre readout side ($x$-axis) at the middle width position. The scan measurement is exemplified in Figure~\ref{fig:PENscan} plotting the light response as a function of the $^{90}$Sr source position. Each point results from averaging 30 multimeter measurements and subtracting the noise value measured with the radioactive source off the scintillator. The noise value is updated at each 5 scan points, as displayed in the bottom pad of the figure. The light response decreases with larger $x$, 
to which corresponds an increasing distance between the origin of the scintillation light  and the collection fibre location (lower $x$), and increasing attenuation effects. The PEN light response is smaller than that of the reference scintillator, but a direct comparison is non-trivial due to the very different geometry. In particular, the thickness difference (2~mm in the PEN sample and 3~mm in the reference scintillator) means a higher energy deposit in the reference scintillator, which by itself would lead to a large number of scintillation photons for the same material type. 

Two parameters of interest are taken from the scans of the different manufactured samples: the maximum light response of the test sample and the light response of the reference scintillator at the point closest to its readout fibre. The maximum light response of the sample is normalised to the reference scintillator measurement to cancel run-to-run fluctuations of the setup response and is used for a  relative comparison between different composition samples. 
The total uncertainty on the normalised light response was found to be 9\%. This corresponds to a standard deviation of the values obtained by measuring the same PEN sample every day throughout the data acquisition weeks. This accounts for uncertainties in the signal measurement, noise instability and the dominant effects of geometric reproducibility resulting from replacing the test samples.

\begin{figure}[t]
\centering
\includegraphics[width=.585\textwidth]{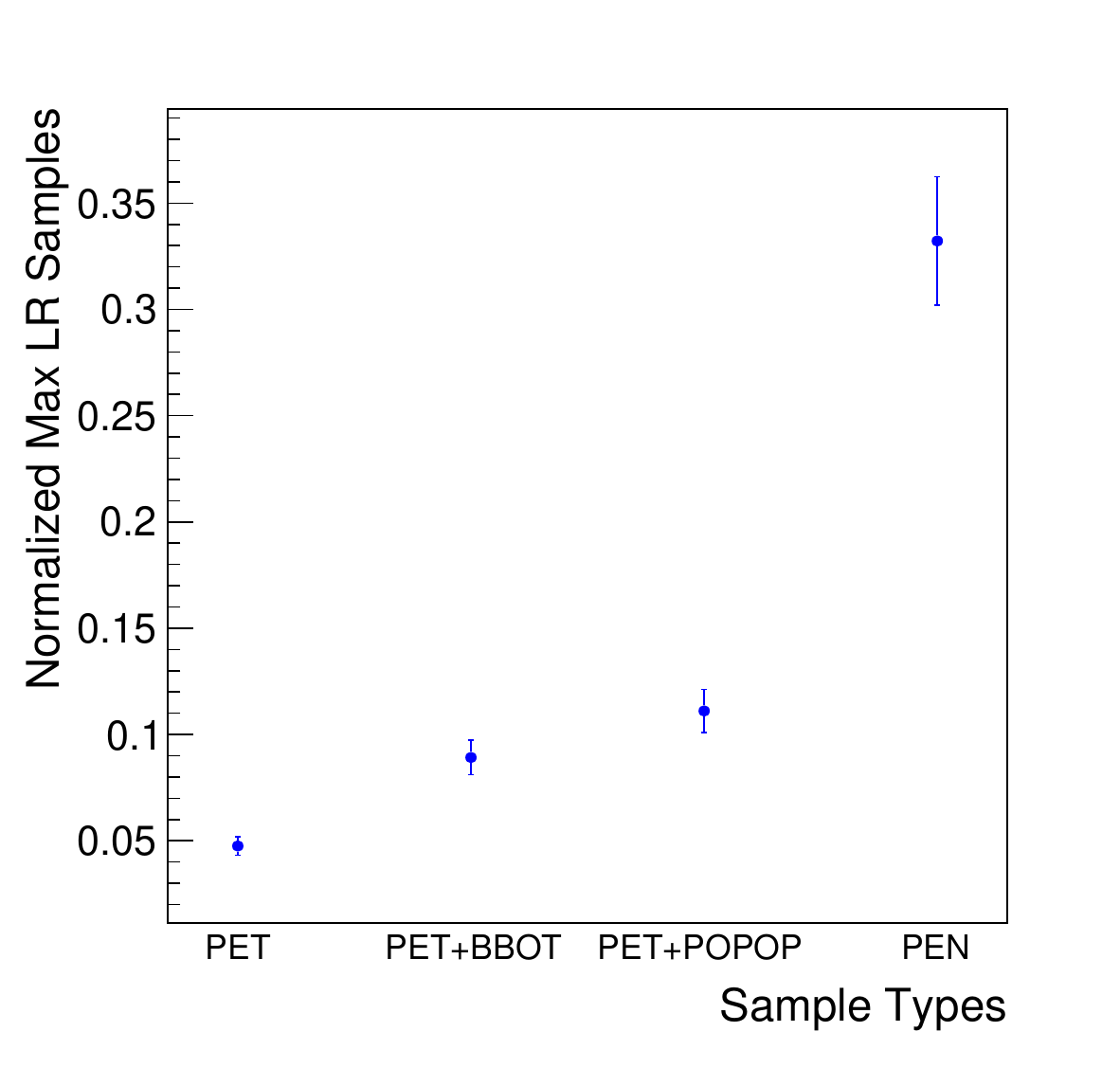}
\caption{Average normalised maximum light response of the PET, PET+BBOT(0.22\%), PET+POPOP(0.22\%) and PEN samples. The vertical error bars correspond the total measurement uncertainty of 9\%.
\label{fig:LRdopants}}
\end{figure}

\begin{figure}[t]
\centering
\includegraphics[width=.585\textwidth]{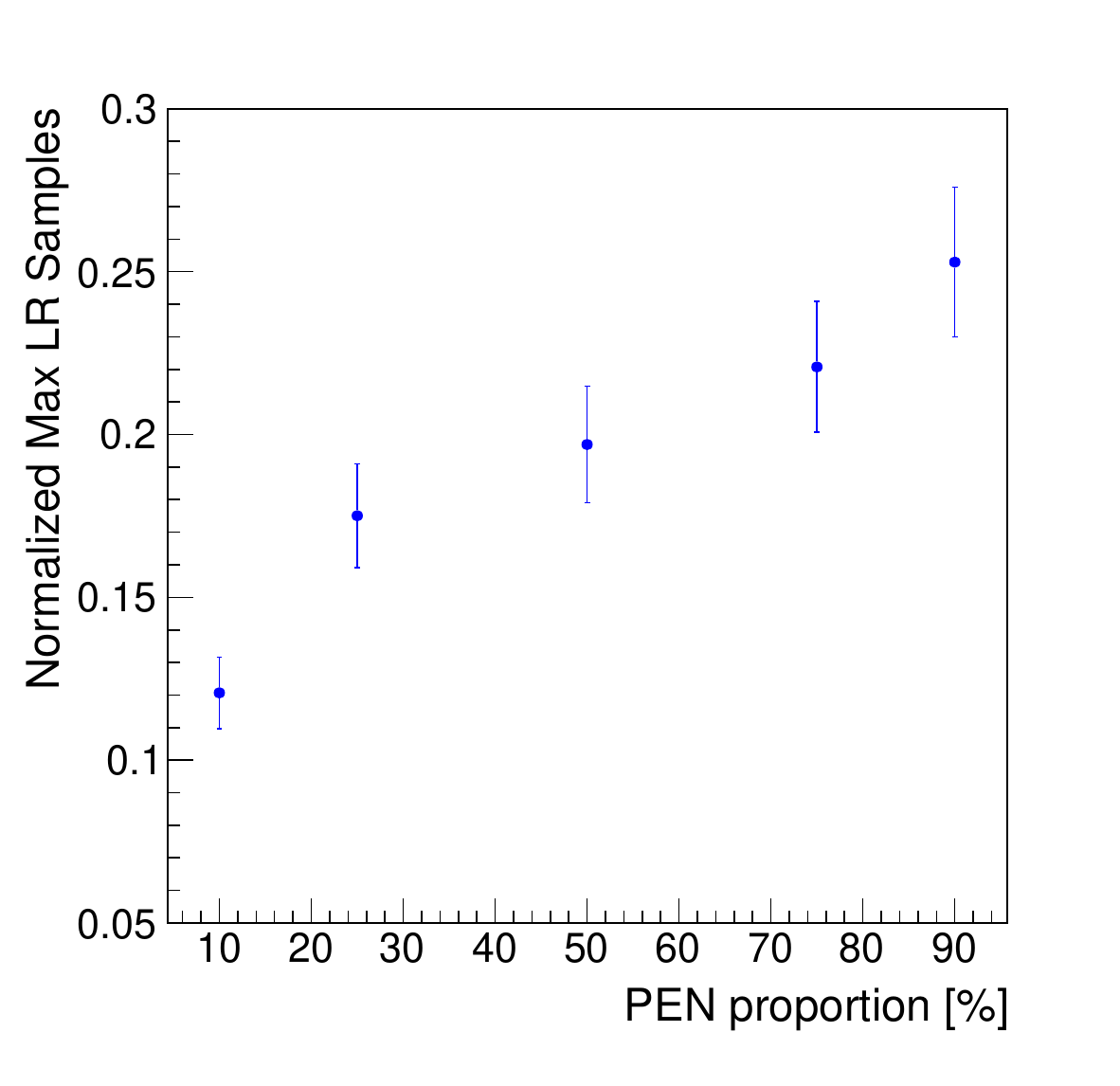}
\caption{Average normalised maximum light response of the PET:PEN mixture samples as a function of the PEN proportion. The vertical error bars correspond the total measurement uncertainty of 9\%.
\label{fig:LRmixturesVsPENproportion}}
\end{figure}

The average normalised maximum light response is plotted in Figure~\ref{fig:LRdopants} for PET, PEN and doped PET samples. The measured light response of PEN is around 7 times larger than that of PET, a factor slightly above the expected from the ratio of absolute light yield ($10500$ and $2200$ photons/MeV for PEN and PET, respectively)~\cite{Nakamura:2011}. This might be due to the differences in the raw material, in the manufacturing process and a better match of the PEN emission spectrum with the Y11 WLS fibre absorption spectrum.
As expected, the addition of dopants to the PET-base material potentiates the light emission. An 80\% and 120\% increase in the light response of PET is obtained when adding 0.22\% mass concentration of BBOT and POPOP, respectively.

The same quantity is plotted for the different PET:PEN mixtures in Figure~\ref{fig:LRmixturesVsPENproportion} as a function of the PEN proportion. Since the homogenisation of the mixtures required a 5-minute staging of the material at processing temperature, which from first principles induces material degradation, the pure PET and PEN samples, not subjected to these conditions, are not plotted for consistency. Relative comparison between different PET:PEN blend proportions is thus obtained, although an upper limit of around 25\% on the light response degradation (due to the 5 minutes staging) could be extracted through comparison with the pure PEN results in Figure~\ref{fig:LRdopants}. The results show an increase of light response with the proportion of PEN, expected given the larger light yield of PEN relative to PET. The trend is compatible with a simple linear dependence, for which case the PET:PEN 25:75 point could be seen as an experimental outlier, or with the typical behaviour of a binary solution, with rapid growth at low concentrations and saturation towards higher values~\cite{Birks}. If the latter is confirmed, it means that a substantial amount of PET could be present in the mixture without significant compromise of light yield but with potential benefit from the PET fast light signals and radiation damage recovery. 

Furthermore, we observe that the addition of POPOP to the PET:PEN mixtures (10:90, 25:75 and 50:50) increase significantly their light response (20 to 40\%), contrarily to BBOT which does not produce the same effect. Despite the visible low quality of the obtained samples, this constitutes evidence of the possible improvement of PET:PEN blends through adequate fluor loading.

\FloatBarrier
\section{Conclusions}

In this paper, we reported the manufacture of $30\times30\times2$~mm$^3$ scintillator samples by injecting moulding PET and PEN commercial grades. For the first time, PET and PEN blends were processed in the laboratory for scintillation applications, and the addition of dopants to pure PET and blended PET:PEN base substrates were explored.
The scintillation properties of the resulting samples were measured. The samples show emission spectra according to the best literature knowledge. 
From the comparative analysis of the light response, we conclude that the light collected by our setup is a factor seven larger for the PEN samples than for the PET samples, and that the addition of BBOT and POPOP in a 0.22\% mass concentration doubles the PET light yield. In addition, we observe a positive correlation between the light response and PEN proportion for PET:PEN blended samples. Finally, we have found first evidence that suitable fluors potentiate the scintillation effect of PET:PEN base mixtures.

\acknowledgments

This work was financed by Fundação para a Ciência e a Tecnologia (FCT), Portugal, through the project EXPL/EME-NUC/1311/2021 - "DLight: New Plastic Scintillators for Future Light-based Detectors"~\cite{DLight}. R. Machado was supported by the PhD fellowship SFRH/PRT/BD/151543/2021 from PT-CERN/IDPASC. B. Pereira was supported by the PhD fellowship SFRH/PRT/BD/152223/2021 from PT-CERN/IDPASC and by the CERN Doctoral Student Programme. R. Pedro was supported by the Junior Researcher fellowship 2021.01023.CEECIND from the Individual Call to Scientific Employment Stimulus by FCT Portugal. The authors thank the Institute of Astrophysics and Space Sciences/FCUL for the equipment and collaboration in the emission spectra measurement.


\bibliographystyle{JHEP}
\bibliography{biblio.bib}

\end{document}